\documentstyle[12pt]{article}
\begin{document}
\thispagestyle{empty}
\def\cqkern#1#2#3{\copy255 \kern-#1\wd255 \vrule height #2\ht255 depth 
   #3\ht255 \kern#1\wd255}
\def\cqchoice#1#2#3#4{\mathchoice%
   {\setbox255\hbox{$\rm\displaystyle #1$}\cqkern{#2}{#3}{#4}}%
   {\setbox255\hbox{$\rm\textstyle #1$}\cqkern{#2}{#3}{#4}}%
   {\setbox255\hbox{$\rm\scriptstyle #1$}\cqkern{#2}{#3}{#4}}%
   {\setbox255\hbox{$\rm\scriptscriptstyle #1$}\cqkern{#2}{#3}{#4}}}
\def\CC{\mathord{\cqchoice{C}{0.65}{0.95}{-0.1}}}
\def\x{\stackrel{\otimes}{,}}
\def\y{\stackrel{\circ}{\scriptstyle\circ}}
\def\proof{\noindent Proof. \hfill \break}
\def\a{\begin{eqnarray}}
\def\b{\end{eqnarray}}
\def\p{{1\over{2\pi i}}}
\def\Q{{\scriptstyle Q}}
\def\P{{\scriptstyle P}}
\renewcommand{\thefootnote}{\fnsymbol{footnote}}

\newpage
\centerline{\LARGE Super--Toda Models Associated}
\centerline{\LARGE to Any (super--)Lie  Algebra}
\vspace{1truecm} \vskip0.5cm

\centerline{\large F. Toppan}
\vskip.5cm
\centerline{Dipartimento di Fisica}
\centerline{Universit\`{a} di Padova}
\centerline{Via Marzolo 8, I-35131 Padova}
\centerline{\em E-Mail: toppan@mvxpd5.pd.infn.it}
\vskip1.5cm
\centerline{\bf Abstract}
\vskip.5cm 
It is shown how to obtain superconformal Toda models  
as reductions of WZNW theories based on any
Lie or super--Lie algebra.
~\par~\par
\pagestyle{plain}
\renewcommand{\thefootnote}{\arabic{footnote}}
\setcounter{footnote}{0}

{\bf Introduction}
\indent

Two-dimensional integrable theories are now
widely studied in the high energy physicists' community
due to their applications to string theory. \par
Two interrelated kinds of such theories are basically considered,
namely the non-relativistic integrable equations in $1+1$ dimensions
(KdV-type systems) and the $2$-dimensional relativistic equations
of Toda type, whose simplest example is the Lioville theory, and 
which can
be regarded as Hamiltonian-constrained WZNW models \cite{ora}.\par
For what concerns the supersymmetric extension of such models it
was known that supersymmetric integrable systems could be obtained, 
in both 
cases, from super--Lie algebras of a particular kind, i.e.  the
class of super--Lie algebras admitting a Dynkin-diagram presentation 
which involves fermionic simple roots only \cite{oi}, \cite{ik}.\par
Later it was understood \cite{das}, \cite{t1} that in the case
of supersymmetric integrable equations of KdV (better to say of 
NLS)-type such requirement was too stringent. A general prescription
to realize non-relativistic
supersymmetric integrable systems from any bosonic or 
super Lie algebra was given in \cite{t2}. \par
In collaboration with D. Sorokin we have further investigated the 
situation for Toda models \cite{st1} and \cite{st2} and realized
that there exists a nice algebraic setting which allows to
construct superconformal Toda theories as WZNW reduction from
any super and bosonic Lie algebra. Here I will present this algebraic
characterization, together with some formulas not published elsewhere.
\par~\par

{\bf The algebraic setting.} 
\indent
Let us recall some basic fact concerning (super-)Lie algebras.\par
Let ${\cal G}$ be a given (super-)Lie algebra having rank $r$.
In a convenient basis it can be reconstructed through
the following set of relations (together with the Serre's relations 
which for our purposes we do not need to specify):
\begin{eqnarray}
\relax [h_i, h_j] &=& 0\nonumber\\
\relax [h_i, e_j] &=& a_{ij} e_j\nonumber\\
\relax [h_i, f_j] &=& - a_{ij} f_j \nonumber\\
\relax [e_i, f_j] &=& \delta_{ij} h_j
\end{eqnarray}
Here $i,j=1,2,...,r$.\\ 
The $h_i$'s are the Cartan generators, while we denote with
$e_i$'s, $f_i$'s 
respectively the positive and negative simple roots. The 
matrix
$a_{ij}$ is the so-called Cartan matrix of ${\cal G}$.\par
Some rank $r=2$ (super-)algebras are
given by $A_2 \equiv sl(3)$ bosonic Lie algebra,
admitting $8$ generators and $2$ invariants having degree 
$2,3$, the $B_2 \equiv sp(4)$ bosonic Lie algebra, with $10$ generators
and $2$ invariants having degree 
$2,4$, the super-Lie algebra $B(0,2)\equiv osp(1|4)$, having $10$ bosonic 
plus $4$ 
fermionic generators (one bosonic and one fermionic simple root).
For completeness let us present the fundamental representations of the 
above (super-)algebras.\\
Let us first denote with $e_{ij}$ the matrices
having entries $c_{ij} = \delta_{ij}$. It turns out that:
\\
the $sl(3)$ algebra is explicitly given by the following $3\times 3$
matrices:
\begin{equation}
\begin{array}{ll}
h_1 = e_{11} - e_{22}; & h_2 = e_{22}- e_{33}\\
e_1 = e_{12}; &  e_2 = e_{23}\\
f_1 = e_{21}; & f_2 = e_{32}
\end{array}
\end{equation}
while the remaining generators associated to the maximal
positive and negative roots are respectively
\begin{equation}
\begin{array}{ll}
e_3 = e_{13}; & f_3 = e_{31}
\end{array}
\end{equation}
Similarly the fundamental representation of the $osp(1|4)$ superalgebra
is realized by the following set of $(4+1) \times (4+1)$ supermatrices
($4$ bosonic and $1$ fermionic index). We have for Cartan generators and 
simple roots:
$$
\begin{array}{ll}
h_1 = e_{11} - e_{22} - e_{33} + e_{44}; & h_2 = e_{33} - e_{44}\\
e_1 = e_{13} + e_{42}; & f_1 = e_{24} + e_{31}\\
e_2 = e_{35} + e_{54}; & f_2 = e_{53} - e_{45}
\end{array} 
$$
($e_1$, $f_1$ are here bosonic while $e_2$, $f_2$ are fermionic).\par
the positive (negative) non-simple roots are given by $p_{kl}$
($n_{kl}$), where $k,l$ label the decomposition in terms of the simple
roots, bosonic and fermionic respectively:
$$
\begin{array}{ll}
p_{02} = 2 e_{34}; & n_{02} = -2 e_{43}\\ 
p_{11} = e_{15} - e_{52}; & n_{11} = - (e_{51} + e_{21})\\
p_{12} = 2 (e_{14} - e_{32}); & n_{12} = 2 (e_{41} - e_{23})\\
p_{22} = - 4 e_{12}; & n_{22} = 4 e_{21} \end{array}
$$
The fundamental representation for the $sp(4)$ subalgebra of $osp(1|4)$ 
is recovered from the above formulas discarding the
fermionic generators $e_2$, $f_2$, $p_{11}$, $n_{11}$. It is therefore
realized in terms of $4\times 4$ bosonic matrices. It should be
noticed that it can be taken
$e_1$ and $p_{02}$ as the two simple positive roots of $sp(4)$.\par
Let us introduce now the differential operator $d$, nilpotent and 
fermionic,
mapping functions into $1$-forms; it turns out 
\a
d^2 &=&0
\label{null}
\b
In the case of one variable $z$, $d = dz {\partial\over\partial z}$. 
Let us denote with $G(z)$ the functions valued in the (super-) 
group $G$, admitting ${\cal G}$ as Lie (super-)algebra.\par
The ${\cal G}$-Lie algebra valued Cartan form 
is introduced through the position 
\a
\Omega =_{def} dG \cdot G^{-1}
\b 
As a consequence of the above definition and the (\ref{null}) property of 
$d$ we get 
the Maurer-Cartan equation
\a
d\Omega - \Omega \cdot \Omega &=& 0
\b
which can also be written, exploiting the Lie-algebraic properties, as 
\a
\relax d\Omega - {\textstyle{1\over 2}} [ \Omega , \Omega]_+ &=& 0
\b
Where the anticommutator is understood in 
the Lie-algebraic context (remember that $\Omega$ is 
a Grassmann $1$-form).\par
The above formula can be trivially extended to the multivariables case,
as well as to the superspace formulation.\par
Let us introduce the $N=1$ superspace with bosonic coordinate $z$ and
fermionic $\theta$. The fermionic $D$ derivative is given by
\a
D&=& {\partial\over \partial \theta} + i\theta \partial_z
\b
Therefore it follows that we can define a nilpotent Grassmann 
differential $d$
\a
d &=_{def}& (dz - i\theta d \theta ) \partial_z + d\theta D
\b
it is easily checked that $d^2=0$ (recall that $dz$ is Grassmann 
but now $d\theta$ is bosonic).
\par
We can introduce as before a Cartan superform which still
satisfies the Maurer-Cartan equation.
Let us denote as $\tau^{\alpha}$ the (super-) generators of the (super-) 
Lie algebra ${\cal G}$. We can therefore set
\a
\Omega &=& (dz - i \theta d \theta ) J + i d\theta \Psi
\b
where $J$, $\Psi$ are ${\cal G}$-valued:  
\a
J= J_{\alpha} \tau^{\alpha} &=_{def}& \partial G \cdot G^{-1}\nonumber\\
\Psi = \Psi_{\alpha}\tau^{\alpha} &=_{def}& -i DG\cdot G^{-1}
\label{comp}
\b
As a consequence of the Maurer-Cartan equation 
satisfied by $\Omega$ the $J_{\alpha}$ superfields are not independent,
but are constructed from the $\Psi_{\alpha}$ superfields:
\a
\relax J &=& D\Psi -{i\over 2} [\Psi , \Psi ]_+
\b
where the anticommutator is in the Lie algebra.\par
It deserves being noticed that $\Psi_{\alpha}$ have opposite statistics, 
while $J_{\alpha}$ have the same statistics of their corresponding 
$\tau^{\alpha}$ generators in ${\cal G}$.\par
To be definite let us take for instance the $A_1\equiv sl(2)$ algebra,
admitting as generators $H, E_+, E_-$ ($\equiv \tau^0, \tau^+,\tau^-
$ respectively) and structure constants given by
\a
\relax [H, E_\pm ] &=& \pm 2 E_\pm
\nonumber\\
\relax [E_+, E_-] &=& H
\b
We obtain\a
J_+ &=& D\Psi_+ - 2 i \Psi_0\Psi_+\nonumber\\
J_- &=& D\Psi_- +2i \Psi_0\Psi_-\nonumber\\
J_0 &=& D\Psi_0 - i \Psi_+\Psi_-
\b
As discussed in our previous paper we can impose on the $N=1$ affine
$sl(2)$ algebra a superconformal 
constraint given by
\a
J_- &=& 1 
\b
which allows us imposing a further gauge-fixing
\a
J_0|_{\theta=0} &=& 0 
\b 
Despite the fact that the above gauge-fixing is not manifestly
supersymmetric it turns out to be indeed superconformal, for details see
\cite{st1}.
\par
The above constraint and gauge-fixing can be explicitly solved in terms 
of the component fields entering the $\Psi_i$ superfields:
Let 
\a
\Psi_i &=& \xi_i (z) + \theta j_i (z)
\b
(here $i=0,\pm)$.
In the $sl(2)$ case we are left with $3$ fundamental unconstrained 
fields, two fermionic and one bosonic,
given by $\xi_-$, $ \xi_+$ and $j_+$, with spin 
dimension respectively $-({1\over 2})$, ${3\over 2}$ and $ 2 $.\par
The remaining fields are expressed through these ones according to:
\a
j_- &=& 1 -i \partial\xi_-\cdot \xi_-\nonumber\\
\xi_0 &=& {1\over 2} \partial \xi_-\nonumber\\
j_0 &=& i \xi_+\xi_-
\b
\par
Let us now discuss the general procedure to impose superconformal
constraints on the $N=1$ affinization of a generic (super)-Lie algebra
${\cal G}$ of rank $r$ with $n_b$ bosonic and $n_f$ fermionic simple
roots (therefore $n_b + n_f = r$). In the following we will take 
letters from the middle of the alphabet, either Latin 
($m,n$) or Greek ($\mu,\nu$), 
to denote respectively bosonic and fermionic simple roots and their
corresponding generators obtained as commutators  
in the Cartan subalgebra (therefore $m,n = 1, 2, ..., n_b$, $\mu,\nu = 
1,2,
... , n_f$).\par
The set of constraints is imposed through the following positions:\\
{\it i}) set $\Psi_a =0$ all superfield associated to every negative
non-simple root\par 
($deg(\tau^a) < -1$).\\
{\it ii~a}) set $\Psi_{-\mu} = 1$ all superfields associated to a fermionic
neagative simple root $\tau^{-\mu}$.\\
{\it ii~b)} set $J_{-m}= 1$ all composite superfields (\ref{comp})
associated to a bosonic negative simple root.\par
The above constraints allows to impose further gauge-fixing 
restrictions which as before turn out to be superconformal. For the 
Cartan
sector we get
\a
\Psi_{0, \mu} &=& 0\quad for\quad any\quad \mu=1,...,n_f \nonumber\\
J_{0, m}|_{\theta =0} &=& 0 \quad for\quad any\quad m=1,..., n_b
\b
Further gauge-fixing conditions can be imposed on the positive-root
sector. We will not specify them in the general case, we will present
them just for the $osp(1|4)$ superalgebra.\par
Due to the constraints and gauge-fixings we have that
\a
J_{-m} &=& D\Psi_{-m} +i \sum_{n=1,...n_b} a_{nm}\Psi_{0,n}\Psi_{-m}
\b
(with $a_{nm}$ elements of the Cartan matrix), 
while
\a
J_{0,m} &=& D\Psi_{0,m} -i \Psi_{+m} \Psi_{-m}
\b
In the particular $osp(1|4)$ case, as already mentioned, 
we deal with one positive
bosonic simple root ($e_1$) and one fermionic ($e_2$). It turns out that
a consistent gauge-fixing condition on the positive sector is given by 
the condition that all superfields $\Psi_{p,kl}$ are set equal to zero
apart $\Psi_{p,02}$, $\Psi_{p,22}$ associated to a $deg =+2, +4$ root 
respectively, as well as $\Psi_{p,12}$ which however is subjected to the
gauge-condition $J_{p,12}=0$ (the corresponding root has degree 
$deg=+3$).\par
The above system of constraints and gauge-fixings can be explicitly 
solved in terms of a set of fundamental unconstrained component fields. 
It turns out that we are left with $5$ surviving fields ($3$ fermionic 
and $2$ bosonic) with spin dimension
given by
\a
 (-{1\over 2},~ {3\over 2},~ 2,~ {5\over 2},~ 4 )&&\nonumber
\b
They are respectively denoted as
\a
(\xi_{f_1},~ \xi_{p,02},~ j_{p,02}, ~\xi_{p,12},~ j_{p,22})&&\nonumber
\b
The remaining field components entering the superfields $\Psi_i$ are 
expressed through the above ones via the relations:
\a
j_{f_1} &=& 1 - i \partial \xi_{f_1}\cdot \xi_{f_1}\nonumber\\
\xi_{h_1} &=& ({1\over 2}) \partial \xi_{f_1}\nonumber\\
j_{p,12} &=& i \xi_{f_1}\partial \xi_{p,12}\nonumber\\
\xi_{p,22} &=& \xi_{f_1} j_{p,22} +({1\over 2}) 
(1 + i\partial \xi_{f_1}\cdot \xi_{f_1} )\partial \xi_{p,12}
\b
All the other component fields are vanishing apart from
\a
\xi_{f_2} &=& 1\nonumber
\b
In the above formulas the meaning of the indices is obvious.\par
As already mentioned such set of constraints and gauge-fixings is 
superconformal. This statement could be proven explicitly 
working at
the purely Lie-algebraic level we have discussed so far. However
the computations, even if straightforward, are rather cumbersome and it is
for that reason more convenient to discuss such topics in the context of
the relation between our Lie-algebraic framework and the Lie-group
properties (this connection is what we need indeed to define super-Toda 
models).
~\par
~\par
~\par
{\bf Super--Toda models.}
\par
Let us come back to the formula which relates the 
superfields
$\Psi_i$ to the functions on the group:
\a
\Psi = \Psi_{\alpha}\tau^{\alpha} &=_{def}& -i DG\cdot G^{-1}
\label{zzz}
\b
According to such a formula the superfield currents in $\Psi$ can be 
expressed in terms of what we can call the Gau\ss~superfields 
entering the Gau\ss~decomposition of the group-valued function $G$:
\a
G&=& M\cdot\Phi\cdot N\nonumber
\b
where $\Phi$ is given by the Cartan ($0$-grading) sector of $G$
\a
\Phi&=& e^{\sum_{i=1,...,r}\Phi_i h_i}\nonumber
\b
and $M$ ($N$) are strictly upper (lower) triangular matrices (we can
assume working in a given, let's say the fundamental, representation for 
the group). Therefore
\a
M&=& {\bf 1} + M_+ + ...\nonumber
\b
and similarly
\a
N&=& {\bf 1} + N_- + ...\nonumber
\b
with $M_+$ ($N_-$) representing the contribution from the $deg = +1$ 
($deg=-1$) elements of the triangular matrices, while the dots represent the 
contributions from the higher (lower) grading elements. 
\par
The WZNW theory can be regarded as a two-dimensional field theory which,
in the supersymmetric case, is determined by the group-valued fields 
$G({Z},{\overline Z})$ which, besides the chiral $N=1$ superspace 
coordinate $Z\equiv, z,\theta$, depends also on the antichiral
superspace coordinate ${\overline Z} \equiv {\overline z}, 
{\overline\theta}$. An antichiral fermionic derivative ${\overline D}$,
\a
{\overline D} &=& 
{\partial\over \partial {\overline\theta}} + i{\overline\theta} 
\partial_{\overline z}
\b
should be introduced. \par
Besides $\Psi$ an antichiral ${\cal G}$-valued superfield ${\overline 
\Psi}$ can be defined
\a
{\overline \Psi} = {\overline\Psi}_{\alpha}\tau^{\alpha} &=_{def}& 
i G^{-1} {\overline D}G
\b
The equations of motion for the unconstrained WZNW model on ${\cal G}$ 
are simply
\a
{\overline D} \Psi = D{\overline \Psi}&=&0
\b
In terms of the component fields the above equation implies that the 
fields entering $\Psi$ are chiral, those entering ${\overline\Psi}$
antichiral.\par
From the above positions and equations of motion
we can extract the constrained WZNZ theory, obtained with the imposition
of the above-discussed superconformal constraints 
and gauge-fixings on the chiral
supercurrents $\Psi$ (and a similar set on the antichiral supercurrents
${\overline\Psi}$).  
\par
As a result the constrained WZNW theory 
turns out to be equivalent to a set of free-field equations
expressing chirality (antichirality) conditions on the set of fields
solving constraints and gauge-fixings, so for instance we get
in the $N=1$ constrained $sl(2)$ case the free equations of motion
\a
{\partial\over\partial z} \xi_- (z) ={\partial\over\partial z} \xi_+(z)=
{\partial\over\partial z } j_+(z)=0 &&
\b
and analogous equations for the antichiral fields.
\par
These free equations are translated 
into non-trivial non-linear equations for the Gau\ss~fields
entering $G$. Let us present here for completeness the relations
between superfields in the unconstrained WZNW model based on $sl(2)$.
\par
In this case we have
\a
G&=& e^{M E_+}e^{\Phi H} e^{N E_-}
\b
and we get
\a
\Psi_+ &=& -i( DM - 2 M D\Phi - M^2 DN e^{-2\Phi})\nonumber\\ 
\Psi_0&=& -i ( D\Phi + M DN e^{-2\Phi})\nonumber\\
\Psi_-&=& -i( DN e^{-2\Phi})
\b
(there are similar equations for the antichiral superfields).
\par
It should be noticed that when discussing the dynamics of the 
constrained supersymmetric WZNW we have a certain freedom in deciding 
which fields should be assumed to be the fundamental ones. We can for 
instance express all the dynamics in terms of the Gau\ss~fields, or
to assume the free currents as the fundamental fields (in this case the
dynamics of the theory gets trivial), but we have also the freedom
to perform a mixed choice, assuming some of the Gau\ss~fields and some of 
the currents as 
fundamental fields. It is indeed convenient to make such a choice as it 
will be clear later. 
\par
Let us discuss now the equations of motion for the constrained WZNW 
models. Let
\a
\Psi &=& \Psi_< + \Psi_= + \Psi_>
\b
where $\Psi_<$, $\Psi_=$, $\Psi_>$ denote the projections
of $\Psi$ onto the negative roots, the Cartan subalgebra and the 
positive roots respectively. Since
\a
\Psi&=& -i ( DM\cdot M^{-1} + M D\Phi \cdot {\Phi}^{-1} M^{-1} + M\Phi
DN \cdot N^{-1}\Phi^{-1}M^{-1})
\b
we get that negative-root terms are obtained only from the third term.
Defined $K$ as
\a
K &=_{def}& -i \Phi D N \cdot N^{-1} {\Phi}^{-1}
\b
and taking into account the constraint on negative 
non-simple roots we get
\a
\Psi_< &=& K
\b
For a generic (super-)Lie algebra we have, due to the constraint 

\a
K&=& \sum_{\mu =1,...,n_f}\tau^{-\mu} + 
\sum_{m=1,...,n_b} \Psi_{-m}\tau^{-m}
\b
The superfields $\Psi_{-m}$ are chiral, but constrained.
In the $osp(1|4)$ case we get for instance
\a
\Psi_< &=& \chi f_2 + f_1
\b
(for simplicity we have set $\chi=_{def} \Psi_{f_2}$).\par
It therefore turns out that
\a
\relax \Psi_= &=&
-i D\Phi \cdot \Phi^{-1} + [ M_+, K] 
\b 
Applying to the above relation the (chiral) equations of
motion for $\Psi$ we get that the Cartan superfields satisfy
\a
\relax -i {\overline D} D( \sum_{i=1,...,r}\Phi_i h_i) 
+ [ {\overline D} M_+ , K]_+
&=& 0
\b
In order to get the equations of motion for the Cartan superfields in
closed form we need to specify ${\overline D} M_+$; this can be done
by looking at the antichiral supercurrents, repeating the same steps as
before. At the end we get, for the $osp(1|4)$ case
\a
{\overline D} M_+ &\propto & e^{\Phi_{h_2}-\Phi_{h_1}} {e_2} +{\overline
\chi} e^{2\Phi_{h_1} -\Phi_{h_2}} e_1
\b
and are led to the following set of equations of motion:
\a
{D}{\chi} + 2 {D} \Phi_1 {\chi}
&=& 1\nonumber\\
{\overline D}{\overline \chi} + 2 {\overline D} \Phi_1 {\overline \chi}
&=& 1\nonumber\\
{\overline D}D\Phi_1 &=&e^{2\Phi_1-\Phi_2}{\overline
\chi}\chi\nonumber\\
{\overline D} D\Phi_2 &=& i e^{\Phi_2-\Phi_1}
\label{eom}
\b
(for simplicity we have set $\Phi_{h_{1,2}}\equiv \Phi_{1,2}$).
A normalization choice, which implies 
in particular the appearance of the $i$ in the r.h.s. has 
been taken in order to guarantee the reality of the superfields
$\Phi_1 , \Phi_2$.\par
It should be noticed that the above set of equations can be considered
as a subset of the equations of motion for the constrained WZNW model.
Indeed this is a closed set of equations involving the Cartan
(i.e. Gau\ss~superfields) $\Phi_{1,2}$ together with the supercurrents
$\chi,{\overline{\chi}}$. However, as it will be clear when performing
the analysis in terms of field components, some extra fields solving
the WZNW constraints, do not enter the above superfields (these are the
spin ${3\over 2}$ fields satisfying the free (chiral and antichiral)
equations. These considerations are particularly important when
performing the hamiltonian derivation of the constrained WZNW system.
Indeed the above equations alone can not be derived from a 
hamiltonian. \par
The reason why some extra fields are present is simply due to the fact 
that in the above derivation we have not used all properties and 
equations in our theory (only the zero-grading component equations
for the Cartan superfields have been taken into account).\par
Before going ahead with the presentation of the above system in terms
of component fields, let us present here their superconformal 
properties.\par
For a single chirality let us consider the infinitesimal local 
transformations
\a
\theta &\mapsto & {\hat\theta} =\theta + \epsilon_f +\theta \partial_z 
\epsilon_b\nonumber\\
z&\mapsto & {\hat z} = z + \epsilon_b + i \theta \epsilon_f
\b
parametrized by the infinitesimal bosonic and fermioni functions
$\epsilon_b,\epsilon_f$ which can be accomodated in the infinitesimal
superfield $\Lambda = i \theta\partial\epsilon_f + \partial\epsilon_b$.
It can be explicitly checked that the equations (\ref{eom}) are 
invariant under the following transformations
\a
\Phi_1 (Z) &\mapsto & {\hat\Phi}_1 ({\hat Z}) = \Phi_1(Z) - 3 \Lambda -
{5\over 2} D\Lambda \chi\nonumber\\
\Phi_2 (Z) &\mapsto & {\hat\Phi}_2 ({\hat Z}) = \Phi_2 - 4 \Lambda 
-{5\over 2} D\Lambda \chi
\nonumber\\
\chi (Z) &\mapsto & {\hat\chi}({\hat Z}) = \chi (Z) + \Lambda 
{\chi}\nonumber\\
D_Z &\mapsto& {\hat D}_{\hat Z} = D_Z - \Lambda D_Z
\b
In  component fields we have
\a
\chi (Z) &=& \xi(z) + \theta j (z)\nonumber\\
{\overline\chi} ({\overline Z}) &=& {\overline \xi}({\overline z})
+{\overline\theta}{\overline j}({\overline z})  \nonumber\\
\Phi_k( Z,{\overline Z}) &=& \phi_k + i\theta \psi_k + 
i{\overline\theta} {\overline \psi}_k + i\theta{\overline\theta} 
B_k\quad for\quad k=1,2
\b
where $\xi,{\overline \xi},\psi_k,{\overline\psi}_k$ are fermionic,
$j,{\overline j}, \phi_k, B_k $ bosonic; moreover all of
them are real.\par 
Due to the equations of motion it turns out that $j,{\overline j}, 
\psi_1,{\overline\psi}_1$ and the auxiliary fields $B_k$ are 
algebraically determined from the remaining fields and their 
derivatives:
\a
j&=& 1-i\partial\xi\cdot\xi\nonumber\\
{\overline j} &=& 1-i{\overline 
\partial}{\overline\xi}\cdot{\overline\xi}\nonumber\\
\psi_1 &=& ({\textstyle{1\over 2}})\partial\xi + \partial\phi_1\cdot\xi\nonumber\\
{\overline\psi}_1 &=& ({\textstyle{1\over 2}}) 
{\overline\partial}{\overline\xi}
+{\overline\partial}\phi_1{\overline\xi}\nonumber\\
B_1 &=& -i e^{2\phi_1-\phi_2}{\overline\xi}\xi\nonumber\\
B_2 &=& e^{\phi_2-\phi_1}
\b
The closed system of equations is given by the following set:
\a
{\overline\partial}\xi &=& 0\nonumber\\
{\partial{\overline\xi}} &=& 0\nonumber\\
{\overline\partial}\psi_2 &=& e^{\phi_2-\phi_1} ({\overline\psi}_2
-({\textstyle{1\over 2}}) {\overline\partial}{\overline\xi} 
-{\overline\partial}\phi_1\cdot {\overline\xi})\nonumber\\
\partial{\overline\psi}_2 &=& -e^{\phi_2-\phi_1} (
\psi_2 -({\textstyle{1\over 2}}) \partial\xi -\partial\phi_1\cdot \xi )
\nonumber\\
\Box \phi_1 &=& -i e^{\phi_1}{\overline \xi}\xi -e^{2\phi_1-\phi_2}(
1-i\psi_2\xi)(1-i{\overline\psi}_2{\overline\xi})\nonumber\\
\Box\phi_2 &=& -e^{2(\phi_2-\phi_1)} + i e^{\phi_1}{\overline\xi}\xi
+ i e^{\phi_2-\phi_1} (\psi_2 -({\textstyle{1\over 2}}) \partial\xi -
\partial\phi_1\cdot\xi ) ({\overline \psi}_2 - ({\textstyle{1\over 2}}) 
{\overline\partial}{\overline\xi} 
-{\overline\partial}\phi_1\cdot{\overline\xi})\nonumber\\
&&
\b
Concerning the above equations some remarks are in order: the system as
already discussed is superconformal. Due to the non-linear 
transformation properties the supersymmetry is spontaneously broken.
Setting $\xi={\overline\xi}=0$ implies reducing the supersymmetric 
system to a system based on standard (not superfields) fields valued
on $osp(1|4)$ which does not admit supersymmetry. Introducing 
the $\xi,{\overline\xi}$ fields allows constructed an enlarged 
supersymmetric system. \par
When setting all fermionic fields 
$\psi_2,{\overline\psi}_2,\xi,{\overline\xi}$ equal to zero we are led
with a coupled system of bosonic equations for $\phi_1,\phi_2$ which
is nothing else than the Toda model associated to the $sp(4)$ bosonic
subalgebra of $osp(1|4)$.\par
Moreover since $osp(1|4)$ admits the superalgebra $osp(1|2)$ as 
subalgebra(associated to the standard superLiouville theory) 
we can ask under which limit the superLiouville theory can be recovered 
from our system. This is just obtained as follows: at first 
one has to set $\xi,{\overline\xi} =0$, and $\phi=_{def}\phi_2-\phi_1$.
Realizing that the limit $\phi_1\rightarrow -\infty$ is a solution of the 
above equations (with the conditions $\xi={\overline\xi}=0$),
then the closed system involving $\phi,\psi_2,{\overline\psi}_2$ is 
nothing
else
than the superLiouville theory (it should be noticed that $\phi_1$ 
should decouple in this case because it is associated to the bosonic 
simple root which does not belong to the $osp(1|2)$ subalgebra of 
$osp(1|4)$).\par
~\par~\par
{\bf Acknowledgements:} It is a pleasure for me to thank D. Sorokin,
the work here presented is based on a collaboration with him. I wish
as well to thank the organizers of the "Semestre sur 
l'integrabilit{\'e}" and the Institut Henri Poincar{\'e} for their
hospitality in Paris, where this work was partly written. 
\par
~\par

\end{document}